\documentclass[11pt]{article}

\usepackage{amssymb}
\usepackage{amsmath}

\def\Dbol{{\stackrel{\circ}{\mathcal D}}{}}

\begin{document}
\renewcommand{\thefootnote}{\fnsymbol{footnote}}
\noindent
{\Large \bf Some Remarks on the Coupling Prescription \\ of Teleparallel Gravity}
\vskip 1.0cm
\noindent
{\bf R. A. Mosna\footnote{Instituto de F\'{\i}sica Gleb Wataghin,
Universidade Estadual de Campinas, 13083-970, Campinas SP, Brazil.
Also at Departamento de Matem\'{a}tica Aplicada, Universidade Estadual de
Campinas, Campinas SP, Brazil.\\ E-mail: mosna@ifi.unicamp.br} \ and
J. G. Pereira\footnote{Instituto de F\'{\i}sica Te\'orica,
Universidade Estadual Paulista, Rua Pamplona 145, 01405-900 S\~ao Paulo SP, Brazil.\\
E-mail: jpereira@ift.unesp.br}
}

\vskip 2.0cm

\begin{abstract}
\noindent
By using a nonholonomic moving frame version of the general covariance principle, an
{\em  active} version of the equivalence principle, an analysis of the gravitational
coupling prescription of teleparallel gravity is made. It is shown that the
coupling prescription determined by this principle is always equivalent with the
corresponding prescription of general relativity, even in the presence of fermions.
An application to the case of a Dirac spinor is made.\smallskip
\end{abstract}
\vskip 1.0cm


\newpage

\section{Introduction: The principle of general covariance}
\renewcommand{\thefootnote}{\arabic{footnote}}
\setcounter{footnote}{0}

The principle of equivalence rests on the equality of inertial and gravitational
masses. It establishes the {\em local} equivalence between gravitational and
inertial effects on all physical systems. An alternative version of this principle
is the so-called {\em principle of general covariance} \cite{sciama}. It states
basically that a physical equation will hold in a gravitational field if it is
generally covariant, that is, if it preserves its form under a general transformation
$x \rightarrow x^\prime$ of the spacetime coordinates. Of course, in the absence
of gravitation, it must agree with the corresponding law of special relativity. The
first statement can be considered as the {\em active} part of the principle in the
sense that, by making a special relativity equation covariant, it is possible to
obtain its form in the presence of gravitation. The second statement can be
interpreted as its {\em passive} part in the sense that the special relativity
equation must be recovered in the absence of gravitation. It is important to
notice that, as is well known, any physical equation can be made covariant through a
transformation to an arbitrary coordinate system. What the general covariance
principle states is that, due to its general covariance, this physical equation
will be true in a gravitational field if it is true in the absence of gravitation
\cite{weinberg}. In other words, to get a physical equation that holds in the
presence of gravitation, the active and the passive parts of the principle must be
true.

In order to make an equation generally covariant, new ingredients are necessary:
A metric tensor and a connection, which are in principle {\em inertial} properties
of the coordinate system under consideration. Then, by using the equivalence
between inertial and gravitational effects, instead of inertial properties, these
quantities can be assumed to represent a {\em true gravitational field}. In this
way,  equations valid in the presence of a gravitational field are obtained from
the corresponding free equations. This is the reason why the general covariance
principle can be considered as an {\em active} version of the {\em passive}
equivalence principle. In fact, whereas the former says how, starting from a
special relativity equation, to obtain the corresponding equation valid in the
presence of gravitation, the latter deals with the reverse argument, that is, it
says that in any {\em locally  inertial} coordinate system, the equations
valid in the presence of gravitation must reduce to the corresponding equations
valid in special relativity.

The above description of the general covariance principle refers to its usual
holonomic version. An alternative, more general version of the principle can be
obtained in the context of nonholonomic moving frames. The basic difference
between these two versions is that, instead of requiring that an equation be
covariant under a general transformation of the spacetime coordinates, in the
moving frame version the equation is required to preserve its form under a {\em
local} Lorentz rotation of the frame. Of course, in spite of the different nature
of the involved transformations, the physical content of both approaches are the
same~\cite{lorentz}.

It is important to emphasize that the principle of general covariance is not an
invariance principle, but simply a statement about the effects of gravitation.
However, when use is made of the equivalence between inertial and gravitational
effects, the principle is seen to naturally yield a gravitational coupling
prescription. By using a moving frame version of this principle, the basic
purpose of this paper will then be to determine the form of the coupling
prescription of teleparallel gravity\footnote{The name teleparallel gravity
is normally used to denote the general three-parameter theory introduced in
\cite{HS}. Here, however, we use it as a synonymous for the teleparallel
equivalent of general relativity, which is the theory obtained for a specific
choice of these parameters.} {\em implied by the general covariance principle}.

\section{Moving frames and associated structures}

We begin by introducing in this section the strictly necessary
concepts associated with moving frames. Let $M$ be a 4-dimensional
Lorentzian manifold representing our physical spacetime. We assume
that $M$ admits a global orthonormal moving frame\footnote{We note
that a classical result \cite{geroch} asserts that every
noncompact spacetime on which spinors may be defined carries a
global orthonormal moving frame.} (or tetrad)
$\{e_{a}\}_{a=0}^{3}$. Let $g$ be the metric on $M$ according to
which the elements of $\{e_{a}\}$ are orthonormal vector fields,
i.e., $g_{x}(e_{a}|_{x},e_{b}|_{x}) = \eta_{ab}$ for each $x \in
M$, with $(\eta_{ab}) = {\rm diag} (1,-1,-1,-1)$. Let
$\{x^{\mu}\}$ be local coordinates\footnote{We use the Greek
alphabet $\mu, \nu, \rho, \dots = 0, 1, 2, 3$ to denote holonomic
spacetime indices, and the Latin alphabet $a, b, c, \dots = 0, 1,
2, 3$ to denote anholonomic indices related to the tangent
Minkowski spaces.} in an open set $U\subset M$. Denoting
$\partial_{\mu}=\partial/\partial x^{\mu}$, one can always expand
the coordinate basis $\{\partial_{\mu}\}$ in terms of $\{e_{a}\}$,
\[
\partial_{\mu} = h^{a}{}_{\mu} e_{a}
\]
for certain functions $h^{a}{}_{\mu}$ on $U$. This immediately yields
$g_{\mu\nu} := g(\partial_{\mu},\partial_{\nu}) = h^{a}{}_{\mu}h^{b}{}_{\nu}
\eta_{ab}$. Let us then see how the global basis of vector fields $\{e_{a}\}$
gives rise to both a Riemannian and a teleparallel structure on $M$.

\subsection{Riemannian structure}

This is obtained by noting that the metric $g$ on $M$ defines a unique
metric-compatible torsion-free connection, which we denote by
$\overset{\circ}{\nabla}$. This is the so-called Levi-Civita connection on
$(M,g)$. It has a possibly nonvanishing curvature
\[
\overset{\circ}{R}(X,Y) e_{a} = (\overset{\circ}{\nabla}_{X}
\overset{\circ}{\nabla}_{Y} -
\overset{\circ}{\nabla}_{X}\overset{\circ}{\nabla}_{Y} -
\overset{\circ}{\nabla}_{[X,Y]}) e_{a},
\]
and its torsion vanishes identically:
\[
\stackrel{\circ}{T}(X,Y) = {\stackrel{\circ}{\nabla}}_{X} Y -
{\stackrel{\circ}{\nabla}}_{Y} X - [X,Y] \equiv 0.
\]
We see in this way that a tetrad can be
used to define a Riemannian structure on $M$.

\subsection{Teleparallel structure}

The moving frame $\{e_{a}\}$ gives rise also to a global notion of parallelism on
$M$. Given two vectors $v\in T_{x}M$ and $w\in T_{y}M$ (with possibly $x\neq y$),
one can simply compare their components with respect to the global frame
$\{e_{a}\}$. This concept can be elegantly formalized by defining another
connection on $M$, according to which the basis vectors $e_{a}$ are parallel. It
is easily seen that there exists a unique connection $\overset{w}{\nabla}$ on $M$
satisfying $\overset{w}{\nabla} e_{a} = 0$. This is the so-called Weitzenb\"{o}ck
connection associated with the moving frame $\{e_{a}\}$ (we note, however, that each
tetrad induces its own $\overset{w}{\nabla}$). Given a vector field $X=X^{a}e_{a}$
on
$M$, we have $\overset{w}{\nabla}_{\mu} X = (\partial_{\mu} X^{a})e_{a}
+X^{a}\overset{w}{\nabla}_{\mu} e_{a} = (\partial_{\mu}X^{a})e_{a}$. Thus, the
$\{e_{a}\}$-components of $\overset{w}{\nabla}_{\mu}X$ are simply the ordinary
derivatives of the $\{e_{a}\}$-components of $X$:
\begin{equation}
\overset{w}{\nabla}_{\mu}(X^{a}e_{a})=(\partial_{\mu}X^{a})e_{a}\text{.}
\label{derivada covariante de Weitzenbock}
\end{equation}
The connection $\overset{w}{\nabla}$ has null curvature $\overset {w}{R}$, but in
general a non-trivial torsion $\overset{w}{T}$. In fact,
\[
\overset{w}{R}(X,Y) e_{a} = (\overset{w}{\nabla}_{X} \overset{w}{\nabla}_{Y}
- \overset{w}{\nabla}_{X}\overset{w}{\nabla}_{Y} - \overset{w}{\nabla}_{[X,Y]})
e_{a} \equiv 0,
\]
for $\overset{w}{\nabla}_{X}e_{a}=0$, $\forall X$. Also, the Weitzenb\"{o}ck
connection yields
\begin{equation}
\overset{w}{T}(e_{a}, e_{b})=-[e_{a}, e_{b}] \equiv - f_{ab}{}^c \, e_c,
\label{torcao e f}
\end{equation}
with $f_{ab}{}^c$ the coefficient of nonholonomy of the basis $\{e_{a}\}$. Notice
that
$f_{ab}{}^c$ can be expressed in terms of the tetrad components $h_a{}^\mu$ by
\begin{equation}
f_{ab}{}^c = (h_a{}^\mu \partial_\mu h_b{}^\nu -
h_b{}^\mu \partial_\mu h_a{}^\nu) h^c{}_\nu.
\label{pcontorsion}
\end{equation}
Now, remember that a local basis $\{e_{a}\}$ of vector fields can be expressed as
a
coordinate basis only if $f_{ab}{}^c=0$. Therefore, the fact that
$\overset{w}{\nabla}$ has a non-null torsion is closely related to the
non-holonomicity of the tetrad $\{e_{a}\}$. It is also important to note that
$\overset{w}{\nabla}$ is compatible with the above defined metric $g$, i.e.
$g_{x}(e_{a}|_{x},e_{b}|_{x})=\eta_{ab}$ for all $x\in M$. In fact, it follows
from Eq.~(\ref{derivada covariante de Weitzenbock}) that $\partial_{\mu
}(g(X,Y))=\partial_{\mu}(X^{a}Y^{b}\eta_{ab}) = g(\overset{w}{\nabla}_{\mu}X,Y)+
g(X,\overset{w}{\nabla}_{\mu}Y)$.

\subsection{Relation between the structures}
\label{sec relation}

As we have seen, each moving frame $\{e_{a}\}$ gives rise to both a Levi-Civita
($R\neq0,$ $T=0$) and a Weitzenb\"{o}ck ($R=0,$ $T\neq0$) connection in $M$. It
is important to keep in mind that, strictly speaking, curvature and torsion are
properties of a connection, and not of spacetime~\cite{livro}. Notice, for
example,
that both the Levi-Civita and the Weitzenb\"ock connections are defined on the
very
same metric spacetime. The difference between these connections defines the
contorsion
tensor ${\stackrel{w}{K}}_{\mu\alpha}{}^{\beta}$ (we recall that the space of
connections is an affine space \cite{KobayashiNomizu}). More precisely, defining
the
connection coefficients $\overset{\circ}{\nabla}{}_{\mu} \partial_{\alpha} =
\overset{\circ}{\Gamma}{}_{\mu\alpha}{}^{\beta}\partial_{\beta}$ and
$\overset{\circ}{\nabla}{}_{\mu}e_{a} = \overset{\circ}{\Gamma}{}_{\mu
a}{}^{b}e_{b}$, with analogous expressions for $\overset{w}{\nabla}$, we have
\begin{equation}
\overset{w}{\Gamma}_{\mu\alpha}{}^{\beta}
= \overset{\circ}{\Gamma}_{\mu\alpha}{}^{\beta}+
{\stackrel{w}{K}}_{\mu\alpha}{}^{\beta},
\label{simbolos de Christoffel e contorcao}
\end{equation}
with
\begin{equation}
{\stackrel{w}{K}}_{abc} = \frac{1}{2}\left({\stackrel{w}{T}}_{cab} +
{\stackrel{w}{T}}_{cba} - {\stackrel{w}{T}}_{bac} \right)
\label{contorsion}
\end{equation}
and $\overset{w}{T}(e_{a}, e_{b})= \overset{w}{T}_{ab}{}^c \, e_c$.
From Eq.~(\ref{simbolos de Christoffel e contorcao}) it follows two important
properties. First, if we choose a local Lorentz frame at a certain point $x\in M$
(free fall), we have that ${\overset{\circ}{\Gamma}}_{\mu\alpha\beta}|_{x} = 0$,
and consequently $\overset{w}{\Gamma}_{\mu\alpha\beta}|_{x} =
{\stackrel{w}{K}}_{\mu\alpha\beta}|_{x}$. Second, seen from the tetrad frame
$\{e_{a}\}$, we have that $\overset{w}{\Gamma}_{\mu ab} \equiv 0$, and thus
$\overset{\circ}{\Gamma}_{\mu ab} = - {\stackrel{w}{K}}_{\mu ab}$. Notice that
this relation holds {\em in this particular frame}. This justifies the apparent
equality  between an affine and a tensor quantities in this expression.

\section{Gravitational coupling prescriptions}

In this section, we use the general covariance principle to study the gravitational
coupling prescription of teleparallel gravity. We start by briefly reviewing the
usual holonomic coordinate version of the principle, as well as the coupling
prescription it implies for the specific case of general relativity.

\subsection{General covariance principle: holonomic coordinate formulation}

Let us consider the Minkowski spacetime (special relativity) endowed with an
\emph{inertial} reference frame with global coordinates $\{x^{\mu}\}$. The motion
of a free particle is then described by
\begin{equation}
\frac{d^{2}x^{\mu}}{ds^{2}}=0.
\label{plivre1}
\end{equation}
In terms of general curvilinear coordinates $\{\bar{x}^{\mu}\}$, the
corresponding equation of motion is given by
\begin{equation}
\frac{d^{2}\bar{x}^{\mu}}{ds^{2}} +
{\bar{\Gamma}}_{\alpha \beta}{}^\mu \,
\frac{d\bar{x}^{\alpha}}{ds} \frac{d\bar{x}^{\beta}}{ds} = 0,
\label{plivre2}
\end{equation}
where the coefficients ${\bar{\Gamma}}_{\alpha \beta}{}^\mu$
are the Christoffel symbols associated with the transformation
$x^{\mu}\rightarrow\bar{x}^{\mu}$. More explicitly,
\[
{\bar{\Gamma}}_{\alpha \beta}{}^\mu =
\frac{1}{2}\bar{\eta}^{\mu\nu}\left(
\frac{\partial\bar{\eta}_{\alpha\nu}}{\partial\bar{x}^{\beta}}+
\frac{\partial\bar{\eta}_{\beta\nu}}{\partial\bar{x}^{\alpha}}-
\frac{\partial\bar{\eta}_{\alpha\beta}}{\partial\bar{x}^{\nu}}
\right),
\]
where $\bar{\eta}_{\mu\nu}$ is the expression of the same \emph{flat} metric, but
now written in terms of the curvilinear coordinates $\bar{x}^{\mu}$, that is,
$\eta = \eta_{\mu\nu} dx^{\mu} dx^{\nu}=
\bar{\eta}_{\mu\nu} d\bar{x}^{\mu} d\bar{x}^{\nu}$.

The general covariance principle can now be invoked to formulate the {\em
physical}
and {\em non-trivial hypothesis} that the metric $\bar{\eta}_{\mu\nu}$ can be
replaced
by a true gravitational field $g_{\mu\nu}$, and consequently
${\bar{\Gamma}}_{\alpha
\beta}{}^\mu$ will represent a dynamical field, given now by
\[
{\stackrel{\circ}{\Gamma}}_{\alpha\beta}{}^{\mu}=\frac{1}{2}g^{\mu\nu}\left(
\frac{\partial g_{\alpha\nu}}{\partial y^{\beta}}+
\frac{\partial g_{\beta\nu}}{\partial y^{\alpha}}-
\frac{\partial g_{\alpha\beta}}{\partial y^{\nu}}
\right),
\]
with its own degrees of freedom. Of course, the metric $g_{\mu\nu}$ reduces to
$\eta_{\mu\nu}$ only when we are in the gravitational vacuum and $y^{\mu}$ are
inertial coordinates. Since ${\stackrel{\circ}{\Gamma}}_{\alpha\beta}{}^{\mu}$
is the symmetric connection, we can say that the general covariance principle
leads naturally to the coupling of General Relativity.

Notice that the minimal coupling prescription is already contained in the above
analysis. To see that, consider a vector field with local expression
$X^{\mu} \partial_\mu$, where $x^{\mu}$ are inertial
coordinates in flat spacetime, as above. Consider the variation
$\partial_{\alpha}X^{\mu}$ of $X^{\mu}$ in the direction given by the
coordinate $x^{\alpha}$. The corresponding expression in curvilinear
coordinates $\{\bar{x}^{\mu}\}$ is easily seen to be
\begin{equation}
D_{\alpha}X^{\mu} := \partial_{\alpha}X^{\mu} +
{\bar{\Gamma}}_{\alpha \beta}{}^\mu X^{\beta}.
\label{covar0}
\end{equation}
Under the hypothesis that in the presence of gravity the corresponding expression
is that obtained by replacing ${\bar{\Gamma}}_{\alpha \beta}{}^\mu$ by
the dynamical field ${\stackrel{\circ}{\Gamma}}_{\alpha\beta}{}^{\mu}$,
we see immediately that
\begin{equation}
\partial_{\alpha}X^{\mu}\rightarrow \Dbol_{\alpha}X^{\mu}:=
\partial_{\alpha}X^{\mu} +
{\stackrel{\circ}{\Gamma}}_{\alpha\beta}{}^{\mu}X^{\beta}
\label{covar1}
\end{equation}
couples gravity to the original field $\partial_{\alpha}X^{\mu}$. This is the well
known minimal coupling prescription of general relativity. The general covariance
principle, therefore, says that gravitation is minimally coupled to matter through
the Levi-Civita connection.

\subsection{General covariance principle: nonholonomic formulation}
\label{sec.moving}

We start again with the special relativity spacetime endowed with the Minkowskian
metric $\eta$.
If $\{x^{\mu}\}$ are inertial Cartesian coordinates in flat spacetime, the basis
of (coordinate) vector fields $\{\partial_\mu\}$ is then a global orthonormal
coordinate basis for the flat spacetime. The frame
$\delta_a=\delta_a{}^\mu\partial_\mu$ can then be thought of as a trivial tetrad,
with components $\delta_a{}^\mu$ (Kronecker delta). Consider now a {\em local}
(that
is, point-dependent) Lorentz transformation
$\Lambda_a{}^b(x)$, yielding the new moving frame
\begin{equation}
e_{a}=e_{a}{}^{\mu }\partial_\mu,
\label{e_a}
\end{equation}
where
\begin{equation}
e_{a}{}^{\mu }(x) = \Lambda _{a}{}^{b}(x)\;\delta _{b}{}^{\mu }.
\label{deltabar}
\end{equation}
Notice that, on account of the locality of the Lorentz transformation,
the new moving frame $e_a = e_{a}{}^\mu \partial_\mu$ is possibly
anholonomous, with
\begin{equation}
[e_a, e_b] = f_{ab}{}^c e_c.
\label{tetradas em espaco chato}
\end{equation}
For this kind of tetrads, defined on flat spacetime, it follows from
Eq.~(\ref{pcontorsion}) that
\begin{equation}
\partial_a (\Lambda_b{}^d)\Lambda^c{}_d =
\frac{1}{2} \left(f^c{}_{ab} + f^c{}_{ba} - f_{ba}{}^c \right),
\label{deltabar e f}
\end{equation}
where $\partial_a=e_a{}^\mu\partial_\mu$ denotes the ordinary directional
derivative along $e_a$, and $\Lambda_a{}^b\Lambda^c{}_b=\delta^c_a$.

The free particle worldline is a curve $\gamma:\mathbb{R\rightarrow}M$, with
$\dot{\gamma} = (dx^{\mu}/ds) \partial_{\mu}$ the particle 4-velocity. In terms of
the moving frame $\{e_{a}\}$, we have $\dot{\gamma} = V^{a} e_{a}$, where
$(dx^{\mu}/ds) = e_a{}^\mu V^{a}$. Seen from the moving frame, a
straightforward calculation shows that the free equation of motion (\ref{plivre1})
can be written in the form
\begin{equation}
\frac{dV^{c}}{ds} + \frac{1}{2}
\left( f^c{}_{ab} + f^c{}_{ba} - f_{ba}{}^c \right) V^a V^b = 0,
\label{plivre5}
\end{equation}
where use has been made of Eq.~(\ref{deltabar e f}). It is important to emphasize
that, although we are in the flat spacetime of  special relativity, we are free to
choose any tetrad $\{e_{a}\}$ as a moving frame. The fact that, for each $x \in M$,
the frame $\{e_{a}|_x\}$ can be arbitrarily rotated introduces the compensating term
$\frac{1}{2}\left( f^c{}_{ab} + f^c{}_{ba} - f_{ba}{}^c \right)$ in the free
particle equation of motion.

Let us now assume that, in the presence of gravity, it is possible to define a
global moving frame $\{e_{a}\}$ on $M$. As we have seen in the preceding section,
such moving frame gives rise to both a Riemannian and a teleparallel structures.
The hypothesis to be made here is that---according to the general covariance
principle---the coefficient of nonholonomy can be assumed to represent a true
gravitational field. In the context of general relativity, as is well known, we
make the identification \cite{mtw}
\begin{equation}
\frac{1}{2} \left( f^c{}_{ab} + f^c{}_{ba} - f_{ba}{}^c \right) =
{\stackrel{\circ}{\Gamma}}_{a b}{}^c,
\label{ricci}
\end{equation}
where ${\stackrel{\circ}{\Gamma}}_{a b}{}^c$ is the Ricci coefficient of rotation,
the torsionless spin connection of general relativity. In this case, the equation of
motion (\ref{plivre5}) becomes
\begin{equation}
\frac{dV^{c}}{ds} + {\stackrel{\circ}{\Gamma}}_{a b}{}^c V^{a} V^{b} = 0,
\label{plivre6}
\end{equation}
which is the {\em geodesic equation} of general relativity. To obtain the equation
of
motion in the teleparallel case, we have to identify, in accordance with
Eq.~(\ref{torcao e f}), the coefficient of anholonomy $f_{ab}{}^{c}$ with {\em
minus} the torsion tensor:
\begin{equation}
f_{ab}{}^{c} = - \stackrel{w}{T}_{ab}{}^{c}.
\label{weitor}
\end{equation}
Accordingly, the equation of motion (\ref{plivre5}) becomes
\begin{equation}
\frac{dV^{c}}{ds} - {\stackrel{w}{K}}_{ab}{}^{c} V^{a} V^{b} = 0,
\label{plivre7}
\end{equation}
where Eq.~(\ref{contorsion}) has been used. This is the {\em force equation} of
teleparallel gravity \cite{glf}. Of course, it is equivalent to the geodesic
equation
(\ref{plivre6}) in the sense that both describe the same physical trajectory.

Now, the above procedure can be employed to obtain a coupling prescription for
gravitation. Consider again a vector field $X$ with local expression $X^{\mu}
\partial_\mu$, where $x^{\mu}$ are inertial coordinates in flat spacetime. Then,
the variation of the components of $X=X^{\mu} \partial_{\mu}$ in the
$\alpha$-direction is trivially given by  $\partial_{\alpha}X^{\mu}$. Still  in
the context of flat spacetime, let us again consider a more general tetrad
$e_{a} = e_{a}{}^{\mu} \partial_{\mu}$ as in Eqs.~(\ref{e_a}) and
(\ref{deltabar}). Let us denote $X=X^{\mu}\partial_{\mu}=X^{a}e_{a},$ where
$X^{a}=e^{a}{}_{\mu}X^{\mu}$. Now, the variation of the components $X^{a}$ must
take into account the intrinsic  variation of $\{e_{a}\}$. A straightforward
calculation shows that $\left[\partial_{\alpha}X^{\mu}\right]\partial_{\mu}=\left[
\partial_\alpha X^c + (\partial_{\alpha} e_b{}^\mu) e^c{}_\mu X^b
\right]  e_{c} = \left[ \partial_\alpha X^c + e^a{}_\alpha
e_{a}(\Lambda_b{}^d)\Lambda^c{}_d X^b
\right]  e_{c}$. It follows from Eq.~(\ref{deltabar e f}) that
\[
\left[\partial_{\alpha}X^{\mu}\right]\partial_{\mu}=
\left[ \partial_\alpha X^c + e^a{}_\alpha
\frac{1}{2}\left( f^c{}_{ab} + f^c{}_{ba} - f_{ba}{}^c \right) X^b
\right]  e_{c}.
\]
Now, by considering again the general covariance principle, we assume that the
presence of gravity is obtained by

(i) replacing $e_a{}^\mu$ by a nontrivial tetrad field $h_a{}^\mu$, which
gives rise to a Riemannian metric $g_{\mu \nu} = \eta_{a b} \, h^a{}_\mu \,
h^b{}_\nu$, and

(ii) replacing the coefficient of anholonomy either by Eq.~(\ref{ricci}) or
(\ref{weitor}). In the first case we obtain
\begin{equation}
\partial_{\alpha}X^{c}\rightarrow {\stackrel{\circ}{D}}_{\alpha} X^{c}:=
\partial_{\alpha}X^{c} + {\stackrel{\circ}{\Gamma}}_{\alpha b}{}^{c}X^{b},
\label{riccimico}
\end{equation}
which is the usual minimal coupling prescription of general relativity. In the
second
case we obtain
\begin{equation}
\partial_{\alpha} X^{c} \rightarrow D_{\alpha} X^{c}:=
\partial_{\alpha} X^{c} - {\stackrel{w}{K}}_{\alpha b}{}^{c} X^{b},
\label{acomivetor}
\end{equation}
which gives the teleparallel coupling prescription of the vector field to gravity
\cite{teleparalelismo}.

The covariant derivative (\ref{acomivetor}) can alternatively be written in the
form
\begin{equation}
D_{\alpha} X^{c} =
\partial_{\alpha} X^{c} - \frac{i}{2} {\stackrel{w}{K}}_{\alpha}{}^{a b}
(S_{ab})^c{}_d X^d,
\label{acomivetor2}
\end{equation}
with \cite{ramond}
\begin{equation}
(S_{ab})^c{}_d = i (\delta_a{}^c \eta_{bd} - \delta_b{}^c \eta_{ad})
\end{equation}
the vector representation of the Lorentz generators. For a field belonging to an
arbitrary representation of the Lorentz group, it assumes the form
\begin{equation}
D_{\alpha} =
\partial_{\alpha} - \frac{i}{2} {\stackrel{w}{K}}_{\alpha}{}^{a b} \Sigma_{ab},
\label{acomivetor3}
\end{equation}
with $\Sigma_{ab}$ denoting a general representation of the Lorentz generators.
This covariant derivative defines the {\em teleparallel} coupling prescription of
fields carrying an arbitrary representation of the Lorentz group.

\section{Dirac spinor field}

\subsection{Dirac equation}

We consider now the specific case of a Dirac spinor \cite{CL}, and apply the general
covariance principle at the level of the Lagrangian formulation.
Once again, let $\{x^{\mu}\}$ be inertial Cartesian coordinates in flat spacetime,
so that $\{\partial_\mu\}$ is an orthonormal coordinate basis for the
flat spacetime. Let $\varphi$ represent a spin-1/2 field
with respect to this frame. The Dirac equation in flat spacetime
can then be obtained from the Lagrangian (we use units in which
$\hbar = c = 1$)
\begin{equation}
\mathcal{L}_M = \frac{i}{2} \Big(
\bar{\varphi}\gamma^a\delta_a{}^\mu \partial_\mu \varphi -
\partial_\mu \bar{\varphi}\gamma^a\delta_a{}^\mu \varphi
\Big) - m \bar{\varphi}\varphi,
\label{lagrDirac0}
\end{equation}
where $\delta_a{}^\mu$ is the trivial tetrad (in the sense that the corresponding
basis of vector fields is just $\{\partial_\mu\}$, as in section \ref{sec.moving}),
$m$ is the particle's mass, $\{\gamma^{a}\}$ are (constant) Dirac matrices in a
given representation, and $\bar{\varphi}=\varphi^{\dagger}\gamma^0$.

Let us consider now, as in section~\ref{sec.moving}, a {\em local} Lorentz
transformation $\Lambda_a{}^b(x)$, yielding the moving frame
$e_a = e_{a}{}^\mu \partial_\mu$ defined by Eqs.~(\ref{e_a})
and (\ref{deltabar}). Substituting $\delta_a{}^\mu=\Lambda^b{}_a \; e_b{}^\mu$,
Eq.~(\ref{lagrDirac0}) reads
\begin{equation}
\mathcal{L}_M = \frac{i}{2} \Big(
\bar{\varphi}\gamma^a \Lambda^b{}_a \partial_b \varphi -
\partial_b \bar{\varphi}\gamma^a\Lambda^b{}_a \varphi
\Big) - m \bar{\varphi}\varphi.
\label{lagrDirac1}
\end{equation}
Taking into account the identity \cite{itzuber}
\begin{equation}
\gamma^{a} \Lambda_a{}^b = L \gamma^b L^{-1},
\label{identity}
\end{equation}
with $L$ a matrix representing an element of the covering group $Spin_{+}(1,3)$
of the restricted Lorentz group, we get
\begin{equation}
\mathcal{L}_M = \frac{i}{2} \left[
\bar{\varphi_e}\gamma^a \left(
\overset{\rightarrow}{\partial}_a - (\partial_a L) L^{-1}\right)\varphi_e -
\bar{\varphi_e} \left(
\overset{\leftarrow}{\partial}_a + (\partial_a L) L^{-1}\right) \gamma^a \varphi_e
\right] - m \bar{\varphi_e}\varphi_e,
\label{lagrDirac2}
\end{equation}
where $\varphi_e = L \varphi$ is the representative of the spinor
field with respect to the moving frame $\{e_a\}$, and $(\partial_a L) L^{-1}$ is a
connection term that appears due to the anholonomicity of the tetrad frame. A
straightforward calculation shows that (see Appendix)
\[
(\partial_{a}L) L^{-1} = - \frac{i}{8} \,
\left(f_{cab} + f_{cba} - f_{bac} \right) \, \sigma^{bc},
\]
with $\frac{\sigma^{bc}}{2}:= \frac{i}{4}[\gamma^{b},\gamma^{c}]$ the spinor
representation of the Lorentz generators. Substituting into Eq.~(\ref{lagrDirac2}),
it becomes
\begin{eqnarray}
\mathcal{L}_M &=& \frac{i}{2} \Big[
\bar{\varphi_e}\gamma^a \Big(
\overset{\rightarrow}{\partial}_a + \frac{i}{8}\left(f_{cab} + f_{cba} - f_{bac}\right)
\sigma^{bc}  \Big)\varphi_e  \nonumber \\
 &-& \bar{\varphi_e} \Big(
\overset{\leftarrow}{\partial}_a - \frac{i}{8}\left(f_{cab} + f_{cba} - f_{bac}\right)
\sigma^{bc}  \Big) \gamma^a \varphi_e
\Big] - m \bar{\varphi_e}\varphi_e. \nonumber
\end{eqnarray}
Now, by considering again the general covariance principle as in section
\ref{sec.moving},
we assume that the presence of gravitation in teleparallel gravity is obtained by
replacing $e_d{}^\mu$ by a nontrivial tetrad field $h_d{}^\mu$, which
gives rise to a Riemannian metric $g_{\mu \nu} = \eta_{a b} \, h^a{}_\mu \,
h^b{}_\nu$, and the coefficient of anholonomy by minus the torsion tensor:
$f_{ab}{}^{c}=-\overset{w}{T}_{ab}{}^{c}$. In addition, the spinor field
$\varphi_e$, as seen from the moving frame $\{e_a\}$, is to be replaced by a
spinor field $\psi$ as seen from the corresponding frame in the presence of
gravitation. This yields the following matter Lagrangian, corresponding to the
Dirac equation for teleparallel gravity
\[
\mathcal{L}_M = \frac{i}{2} \left[
\bar{\psi}\gamma^a \left(
\overset{\rightarrow}{\partial}_a - \frac{i}{4}K_{abc} \sigma^{bc}
\right)\psi - \bar{\psi} \left(
\overset{\leftarrow}{\partial}_a + \frac{i}{4}K_{abc} \sigma^{bc}
\right) \gamma^a \psi
\right] - m \bar{\psi}\psi,
\]
where Eq.~(\ref{contorsion}) has been used. Equivalently, we can write
\begin{equation}
\mathcal{L}_M = \frac{i}{2} \Big(
\bar{\psi} h_a{}^\mu\gamma^a D_\mu \psi -
D_\mu \bar{\psi} h_a{}^\mu\gamma^a \psi
\Big) - m \bar{\psi}\psi,
\label{diraclagr}
\end{equation}
with the teleparallel version of the Fock-Ivanenko derivative operator given
by \cite{tsc}
\begin{subequations}
\label{D teleparalelo}
\begin{align}
D_{\mu}\psi & =
\partial_{\mu}\psi-\frac{i}{4}\;{\overset{w}{K}}_{\mu bc}\,{\sigma^{bc}}\psi,
\label{D teleparalelo a}\\
D_{\mu}\bar{\psi}  & =
\partial_{\mu}\bar{\psi}+\frac{i}{4}\;\bar{\psi}{\overset{w}{K}}_{\mu bc}\,{\sigma^{bc}}.
\label{D teleparalelo b}
\end{align}
\end{subequations}
This gives the coupling prescription for spin-1/2 fields in the teleparallel
formalism. Moreover, in terms of the underlying Riemannian structure,
Eq.~(\ref{D teleparalelo a}) can be rewritten in the form
\[
{\stackrel{\circ}{D}}_{\mu}=\partial_{\mu} +
\frac{i}{4} \; {\overset{\circ}{\Gamma}}_{\mu ab} \; \sigma^{bc},
\]
which is the well-known minimal coupling prescription of general relativity, as
defined by the usual Fock-Ivanenko derivative.

A straightforward calculation shows that the matter Lagrangian (\ref{diraclagr})
gives rise to the equation of motion
\begin{equation}
i \gamma^a h_a{}^\mu D_\mu \psi = m\psi,
\label{diracequation}
\end{equation}
which is the Dirac equation in teleparallel gravity. It is interesting to note
that Eq.~(\ref{diracequation}) can also be obtained directly from the flat
spacetime Dirac equation
\[
i \gamma^{a} \delta_a{}^\mu \partial_{\mu} \psi= m \psi
\]
by substituting $\partial_\mu$ and $\delta_a{}^\mu$ with $D_\mu$ and
$h_a{}^\mu$. This is a distinguished feature of the coupling  considered
here, since it is well known that the usual minimal coupling prescription
in Riemann-Cartan spacetimes leads to different results when applied to the
Lagrangian or to the field equations \cite{saa}.

\subsection{Irreducible decomposition for torsion}

We decompose now the torsion tensor in irreducible components under the global
Lorentz group,
\[
T_{abc}=\frac{1}{3}(T_{a} \eta_{bc} - T_{b} \eta_{ac}) + q_{abc} -
\frac{1}{6}\epsilon_{abcd}S^{d},
\]
where $T_{a}=T_{ab}{}^{b}$ is the torsion trace, $S^{d}:=\epsilon^{abcd}T_{abc}$
is the torsion pseudo-trace, also called axial torsion, and $q_{abc}$ has null trace
and null pseudo-trace. A straightforward calculation then yields
\[
- \frac{i}{4}K_{abc}\gamma^{a} \sigma^{bc}= \gamma^{a}
\left( \frac{1}{2}T_{a} - \frac{i}{8} S_{a} \gamma^{5}  \right),
\]
where $\gamma^{5}=\gamma_{5}:=i\gamma^{0}\gamma^{1}\gamma^{2}\gamma^{3}$.
After substituting in Eq. (\ref{diracequation}), we get (see also
\cite{ZhangBeesham})
\begin{equation}
\gamma^{\mu}(x)\left(
i\partial_{\mu} + \frac{i}{2}T_{\mu} + \frac{1}{8} S_{\mu}\gamma^{5}
\right)  \psi = m \psi,
\label{eq de Dirac teleparalela irredutivel}
\end{equation}
where $\gamma^{\mu} \equiv \gamma^{\mu}(x) = h_a{}^\mu \gamma^a$.
This is the Dirac equation in the teleparallel formalism in terms of
irreducible components for torsion.

In contrast to the usual General Relativity formalism, where the
gravitational coupling is given by the \emph{affine} quantities
$\overset{\circ}{\Gamma}_{\mu a}{}^b$, in the above equation all
coupling terms have a \emph{tensorial} or \emph{pseudo-tensorial}
nature. We observe that this is not a contradiction, since in
the teleparallel formalism a global moving frame $\{e_{a}\}$ is
\emph{fixed}. This fixes, in a sense, the gauge corresponding to
the affine transformations in $\overset{\circ}{\Gamma}_{\mu a}{}^{b}$.
As a result, the Dirac equation exhibits a simpler final form, since
the gravitational coupling is then realized through vector ($T_{\mu}$)
and pseudo-vector ($S_{\mu}$) quantities. It is interesting to notice
that, if $\psi$ is a state with a definite parity (say, $P \psi = + \psi$),
then the axial torsion $S_\mu$ will couple to the state with opposite
parity $\gamma^{5} \psi$.

\section{Final remarks}

The general covariance principle can be considered as an {\em active} version of
the {\em passive} equivalence principle. In fact, whereas the former says how,
starting with a special relativity equation, to obtain the corresponding equation
valid in the presence of gravitation, the latter deals with the reverse argument,
namely, that in a {\em locally inertial} coordinate system any equation of general
relativity must reduce to the corresponding equation of special relativity. More
specifically, what the general covariance principle states is that any physical
equation can be made covariant through a transformation to an arbitrary coordinate
system, and  that, due to its general covariance, this physical equation will be
true in a  gravitational field. Of course, the passive equivalence principle must
also hold. Now, in order to make an equation generally covariant, new ingredients
are necessary: A metric tensor and a connection, which are in principle {\em
inertial} properties of the coordinate system. Then, by using the equivalence
between inertial and gravitational effects, instead of inertial properties, these
quantities can be assumed to represent a {\em true gravitational field}. In this
way, equations valid in the presence of a gravitational field can be obtained.

The above description of the general covariance principle refers to its usual
holonomic version. An alternative, more general version of the principle can be
obtained in the context of nonholonomic moving frames, whose application is
mandatory, for example, in the presence of spinor fields. The basic difference
between these two approaches is that, instead of requiring that an equation be
covariant under a general coordinate transformation, in the moving frame version
the equation is required to preserve its form under a local Lorentz rotation of the
frame. Of course, in spite  of the different nature of the involved transformations,
the physical content of both approaches is the same.

An important property of the general covariance principle is that it naturally
yields a coupling prescription of any field to gravitation. In other words, it
yields the form of the spin connection appearing in the covariant derivative. By
using the nonholonomic version of this principle, we have found that the spin
connection of teleparallel gravity is given by {\it minus} the contorsion tensor, in
which case the coupling prescription of teleparallel gravity becomes always
equivalent to the corresponding prescription of general relativity, even in the
presence of spinor fields.

Finally, we remark that the form of the coupling prescription for spinor fields in
teleparallel gravity has been a matter of recurrent interest \cite{tsc,maa,maluf}
(see also \cite{MV,maareply}). In \cite{maa}, the authors argue that,
when applied to teleparallel gravity, the minimal coupling associated with
the Weitzenb\"{o}ck connection leads to some inconsistencies. As shown in
\cite{maluf}, such problems do not arise if the adopted teleparallel coupling
is equivalent to the corresponding prescription of General Relativity. The
analysis presented here shows that such coupling follows naturally
from the general covariance principle.

\section*{Acknowledgments}
RAM thanks W. A. Rodrigues for useful discussions.
The authors are grateful to FAPESP and CNPq for financial support.

\begin{appendix}
\section*{Appendix}

As we have already mentioned, $L$ is an element of $Spin_{+}(1,3)$, the covering
of the restricted Lorentz group. Thus, $(\partial_{a}L) L^{-1}$ belongs to the
corresponding Lie algebra, which is generated by
$\frac{\sigma^{bc}}{2} = \frac{i}{4}[\gamma^{b},\gamma^{c}]$. Therefore, we
can write
\begin{equation}
(\partial_a L)L^{-1} = \frac{1}{2} \; \omega_{abc} \, \frac{\sigma^{bc}}{2},
\label{eq1a}
\end{equation}
for certain functions $\omega_{abc}$ satisfying
$\omega_{abc}=-\omega_{acb}$. These functions are actually Lorentz-valued
connections representing the gravitational vacuum, that is, a connection
accounting
for the frame anholonomy only.

Let us take now the identity (\ref{identity}). Taking the derivative on both
sides,
\[
(\partial_a L)\gamma^e L^{-1}+L \gamma^e \partial_a L^{-1}=
(\partial_a \Lambda_d{}^e)\gamma^d,
\]
which is equivalent to
\[
\Lambda_d{}^e \left[ (\partial_{a} L) L^{-1},\gamma^{d}\right]  =
(\partial_{a}  \Lambda_d{}^e) \gamma^{d}.
\]
Substituting (\ref{eq1a}), we get
\[
\frac{1}{2} \omega_{abc} \, \Lambda_d{}^e \left[ \frac{\sigma^{bc}}{2}, \gamma^{d}
\right] =(\partial_{a} \Lambda_d{}^e) \gamma^{d}.
\]
Using the commutation relation
$\left[ \frac{\sigma^{bc}}{2},\gamma^{d}\right] = i
(\eta^{cd}\gamma^{b}-\eta^{bd}\gamma^{c}$), we get
\[
\omega_{ab}{}^c = - i (\partial_a \Lambda_b{}^d)\Lambda^c{}_d .
\]
It follows from (\ref{deltabar e f}) that
\[
\omega_{ab}{}^c = - \frac{i}{2} \left(f^c{}_{ab} + f^c{}_{ba} - f_{ba}{}^c
\right),
\]
which finally implies
\begin{equation}
(\partial_{a}L)L^{-1} = - \frac{i}{4}
\left(f_{cab} + f_{cba} - f_{bac} \right) \, \frac{\sigma^{bc}}{2}.
\label{eq3a}
\end{equation}
\end{appendix}


\end{document}